\documentclass[%prl,
aps,%preprint %tightenlines
 ,twocolumn
]{revtex4}
\usepackage{graphicx}

\begin{document}

\title{{\itshape Relativistic quantum information and time machines}}

\author{Timothy C. Ralph and Tony G. Downes\\
School of Mathematics and Physics, University of Queensland, Brisbane, Australia\\ }

\begin{abstract}
Relativistic quantum information combines the informational approach to understanding and using quantum mechanics systems - quantum information - with the relativistic view of the universe. In this introductory review we examine key results to emerge from this  new field of research in physics and discuss future directions. A particularly active area recently has been the question of what happens when quantum systems interact with general relativistic closed timelike curves - effectively time machines. We discuss two different approaches that have been suggested for modelling such situations. It is argued that the approach based on matching the density operator of the quantum state between the future and past most consistently avoids the paradoxes usually associated with time travel.
%\bigskip

%\begin{keywords} Quantum information, relativity, quantum communication, space-time curvature, closed timelike curves
%\end{keywords}\bigskip
%\bigskip

\end{abstract}

\maketitle

\section{Introduction}

Our two most fundamental theories of the universe are general relativity and quantum mechanics. Both exhibit highly non-intuitive features from the point of view of everyday experience, but both work extremely well in predicting and quantifying phenomena in their respective regimes of application. Although special relativity and quantum mechanics have long been unified in principle, there remain many situations in which the consequences of this union are poorly understood, and a full quantum theory of gravity remains elusive even in principle. The non-intuitive features of both theories lead to apparent paradoxes when pushed to extremes. For example in quantum mechanics it is in principle possible to consider a macroscopic object (such as a cat) being in a superposition of macroscopically distinct states (such as dead and alive) - the Schr\"odinger's cat paradox. In general relativity it is mathematically allowed to construct time machines that would allow one to travel into ones own past creating an acausal loop referred to as a closed timelike curve. It remains an open question as to whether such bizarre predictions can actually be realized, and if so how the apparent paradoxes involved would be resolved.

In the last couple of decades significant new insights into quantum mechanics, and new applications for quantum mechanical systems, have arisen from the study of quantum information science. Quantum information is the study of the encoding, transmission and processing of information stored in quantum systems \cite{nielsen:2000}. The unique properties of quantum mechanics leads to quite different rules to those previously discovered in classical information theory. On one hand these new rules lead to the ability to perform information tasks that are hard or impossible for classical systems. More fundamentally, the informational point of view, and in particular the concept of the quantum bit or qubit, has proved a unifying concept across many areas of quantum physics. 

Quantum Information protocols can be broadly divided into: communications; measurement; and computation. Quantum communication protocols exhibit intrinsic security and enhanced channel characteristics over classical communications. Quantum measurement protocols optimise the precision with which parameters characterising quantum systems, or classical systems with which they interact, are extracted. Quantum Computation exhibits faster processing times for a number of key tasks, algorithms and simulations. In certain cases this makes problems that are completely intractable to classical computation solvable. 

Entanglement is a key resource in virtually all quantum information protocols \cite{plenio:1998}. Entanglement refers to the uniquely quantum correlations that can exist between spatially separated quantum sub-systems. A simple example of an entangled state is the singlet state:
\begin{equation}
|\phi^- \rangle = 1/\sqrt{2}(|0 \rangle_A |1 \rangle_B - |1 \rangle_A |0 \rangle_B)
\end{equation}
where $|0 \rangle$ and $|1 \rangle$ are logical qubit states that can be used to represent any quantum 2-level system and the subscripts $A$ and $B$ label spatially separated sub-systems. The state $|\phi^- \rangle$ is said to be entangled because it is not possible to write it in the separable form $|\rho \rangle_A |\sigma \rangle_B$. We may also refer to $|\phi^- \rangle$ as a Bell state, so called because it maximally violates the Bell inequality - a correlation bound that no unentangled state can break. Entropy can be used to quantify the information content of qubits \cite{nielsen:2000}. A state with zero entropy is said to be pure - there is nothing more we can learn about it. States with non-zero entropy are referred to as mixed and are represented by density operators. Although $|\phi^- \rangle$ is a pure state, its subsystems, $\rho_A$ and $\rho_B$, are mixed, where the density operator of one sub-system is obtained by taking the partial trace over the other sub-system, e.g. $\rho_A = Tr_B(|\phi^- \rangle \langle \phi^- |)$. This generic feature of entangled states - that we know more about the total system than we know about its individual parts - has no equivalent in classical information theory.

Most quantum information research to date has been based on non-relativistic quantum mechanics and thus is an approximation. As the distances over which quantum information protocols are carried out steadily increase \cite{fedrizzi:2009}, with quantum operations being carried out on shorter and shorter time-scales, and with plans to place quantum sources into space \cite{ursin:2008} for communications and metrology applications, this approximation will begin to  fail. Already, classical space technologies like the Global Positioning System must account for the slowing of clocks in gravitational fields in order to achieve their current precision \cite{taylor:2000}. %and, in the quantum domain, experiments have been carried out to exploit the finite speed of light in designing stronger quantum communication protocols \cite{kwiat:2006}. 
If information is carried by quantum systems then the abstract rules of information science developed for classical systems must be modified. If we must now consider information carried on relativistic quantum systems then the rules will need to be modified further. The expansion of quantum information science to a fully relativistic setting will modify and perhaps limit existing quantum information protocols. However it is also expected to lead to new protocols and a deeper understanding of the physical universe more generally.

In this paper we review progress in this new field of relativistic quantum information and then particularly look at what the quantum informational view of physics can tell us about the apparent paradoxes of relativistic time travel. In the following section we examine how quantum information is affected by relativistic motion, firstly by inertial observers, and then non-inertial observers. We then discuss the entanglement properties of the Minkowski vacuum (flat-space) before extending our discussion to curved space-times. In section 3 we focus on time machines, firstly as they arise in general relativity, and then time travel interpretations in quantum mechanics. We then examine and contrast two distinct ways to treat qubits interacting with closed timelike curves formed by relativistic time machines. In the final section we conclude and discuss possible future directions.

\section{Relativistic Quantum Information}

\subsection{Inertial Observers}

The effects of relativistic motion on quantum systems was considered early in the development of quantum theory. With the development of quantum information theory the question arises once again, what are the effects of relativistic motion on the concepts developed in quantum information theory? It is clear that different observers may disagree on the value of qubits encoded in a quantum system unless they use a consistent coordinate system. However it is more surprising that quantities such as entanglement between qubits \cite{czachor:1997, gingrich:2002} and purity, and hence entropy, of individual qubits \cite{peres:2002} can vary between different inertial observers , i.e observers traveling with different, constant velocities. 

It is known that for entanglement, the degree of violation of the Bell inequality depends on the velocity of the particles when relativistic effects are taken into account. %\cite{czachor:1997, ahn:2003}. 
The basic mechanism here is an observer dependent coupling between the discrete qubit degree of freedom (for example spin) and the linear momentum degrees of freedom. Lorentz transformations predict that a moving observer sees a rotation of particle spin with respect to that seen by a stationary observer. The magnitude of this spin rotation is a function of the particle's linear momentum. Quantum particles are described by a wave function, comprised of a superposition of many momentum states. Hence the Lorentz transformation induces correlations between spin and linear momentum. The mixing of spin and momentum under Lorentz transformations will subsequently effect the strength of entanglement between two spin systems \cite{gingrich:2002} (see Fig.1). 

The entanglement between the spin degrees of freedom and the momentum degrees of freedom are seen to change between frames. Indeed the relativistic transformation between two different reference frames will effect the degree of entanglement between the spin and momentum degrees of freedom of an individual particle. If one chooses not to measure the momentum degree of freedom this entanglement will effect the reduced density matrix, and hence the purity of the qubit will change for different observers. This appears to demonstrate that even the entropy of a single qubit is not a Lorentz invariant concept \cite{peres:2002}. However, the total purity, taking all degrees of freedom into account, is a Lorentz invariant quantity. 
\begin{figure}[htb]
\begin{center}
\includegraphics*[width=7.5cm]{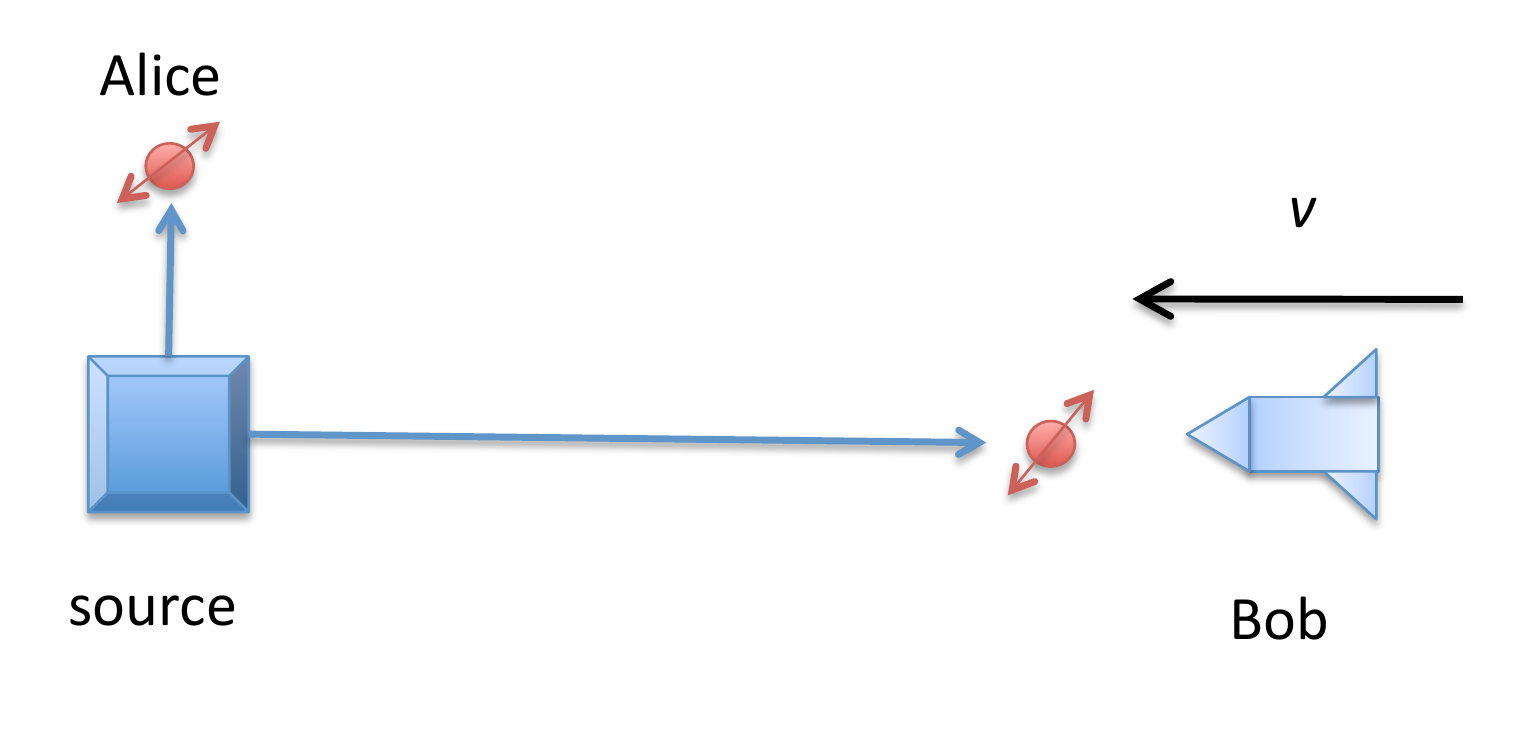}
\caption{Alice holds a source of spin entangled pairs. She keeps one and sends the other to Bob. If Bob is in the same reference frame as Alice ($v=0$) then they find maximal entanglement between their particles. However if Bob is in a different inertial frame ($v \ne 0$) then they find reduced entanglement and conclude the joint state of the spins is mixed. The reduction in entanglement will be approximately proportional to $\Delta v/(m c^2)$, where $\Delta$ is the momentum spread of the wavefunction and $m$ is the mass of the spin. We have assumed $v << c$.
}
\label{embedwormhole}
\end{center}
\end{figure}

These results also apply to qubits other than spin, for example photon polarisation \cite{gingrich:2003}. The effect on photon polarisation was immediately seen to impact quantum cryptographic protocols that require photons to transmit quantum information \cite{czachor:2003}. A solution to the problem was found by replacing the helicity polarisation basis used previously, with a relativistic definition of polarisation. The new construction gives qubits which are consistent in different frames and allow for quantum cryptographic protocols to be carried out between inertial observers. 

These early developments have been reviewed in detail by Peres and Terno \cite{peres:2004}. 
More recent investigations of how Lorentz transformations affect quantum information processes and entanglement have investigated arbitrary spin statistics and quantum cloning  \cite{kok:2006, bradler:2008},
%, moradi:2009b
including a detailed analysis of the entanglement between the various partitions of the momentum and spin degrees of freedom for two massive particles \cite{friis:2010}. What has gained significant attention since the early work has been the possibility of encoding quantum information in ways which are invariant under Lorentz transformations. This has included work on qubits \cite{bartlett:2005} and continuous variables \cite{kok:2005}.% as well as Bell's inequality and entanglement \cite{kim:2005, harshman:2005, caban:2006, caban:2007, lamata:2006} and photonic wave packets \cite{bradler:2010}. 

\subsection{Accelerated observers}

If we consider quantum protocols between non-inertial observers, then an additional effect that should be taken into account is the possibility of particle creation via the Unruh-Davies effect \cite{Unruh:1976, davies:1975}. The Unruh effect occurs because the very notion of particles, like photons and electrons, is an observer-dependent concept \cite{wald:1994}. In flat space-time, one may consider the quantum theory of a field as observed by any inertial observer. One may also consider the theory for the same field, as observed by a uniformly accelerating observer. When these two theories are compared, one finds that they do not match up \cite{fulling:1973}. In particular, in flat space-time there exists a state of the field, the vacuum state, which appears to all inertial observers as completely empty of particles everywhere. The same state will appear to a uniformly accelerating observer as thermalised, with particles at the Unruh temperature given by \cite{crispino:2008}:
\begin{eqnarray}
T=\frac{a\hbar}{2\pi kc}
\label{U}
\end{eqnarray}
Here $a$ is the rate of acceleration of the observer, $\hbar$ is the reduced Planck's constant, $k$ is Boltzmann's constant and $c$ is the speed of light. The fact that these are real particles can be seen when one calculates the response of a uniformly accelerating detector in the Minkowski vacuum. One finds that the probability of the detector becoming excited follows a thermal distribution at the Unruh temperature. The Unruh effect requires very high accelerations - from Eq.\ref{U} one finds $1^{\circ}$K corresponds to $a \sim 10^{20} ms^{-2}$ - and has not yet been experimentally observed.

The thermal nature of the Unruh effect occurs because of two main reasons. Firstly their exists a large amount of entanglement in the vacuum state of a field in flat space-time, as will be described more in section 2.4. Secondly a uniformly accelerating observer will perceive a causal horizon to exist, which prevents any exchange of information with observers on the opposite side of the horizon. This ignorance of the state on the other side of the horizon, together with the quantum correlations which exist between the two regions, leads to an increase in uncertainty as perceived by the uniformly accelerating observer. This uncertainty is manifested as the thermalisation of the particles observed. Another way of saying this is that the entire space-time vacuum is in a pure entangled state, but if we are restricted to observing only a section of that space-time, then that sub-system will be in a mixed state.

Fig.2 shows a space-time diagram with Minkowski coordinates $(t,x)$ which are appropriate for inertial observers and Rindler coordinates $(\eta,\xi)$ which are appropriate for uniformly accelerating observers. 
The dashed lines show the causal horizon experienced by the accelerating observers, who travel along the hyperbolic trajectories.  The two coordinate systems are related within the right hand sector (the right Rindler wedge) via:
\begin{eqnarray}
t &= & a^{-1} e^{a \xi} sinh(a \eta); \;\;\;\;\;\;\; z = a^{-1} e^{a \xi} cosh(a \eta); 
\label{R1}
\end{eqnarray}
and in the left Rindler wedge via:
\begin{eqnarray}
t &= & -a^{-1} e^{a \bar \xi} sinh(a \bar \eta); \;\;\;\;\;\;\; z = -a^{-1} e^{a \bar \xi} cosh(a \bar \eta). 
\label{R2}
\end{eqnarray}
\begin{figure}[htb]
\begin{center}
\includegraphics*[width=7.5cm]{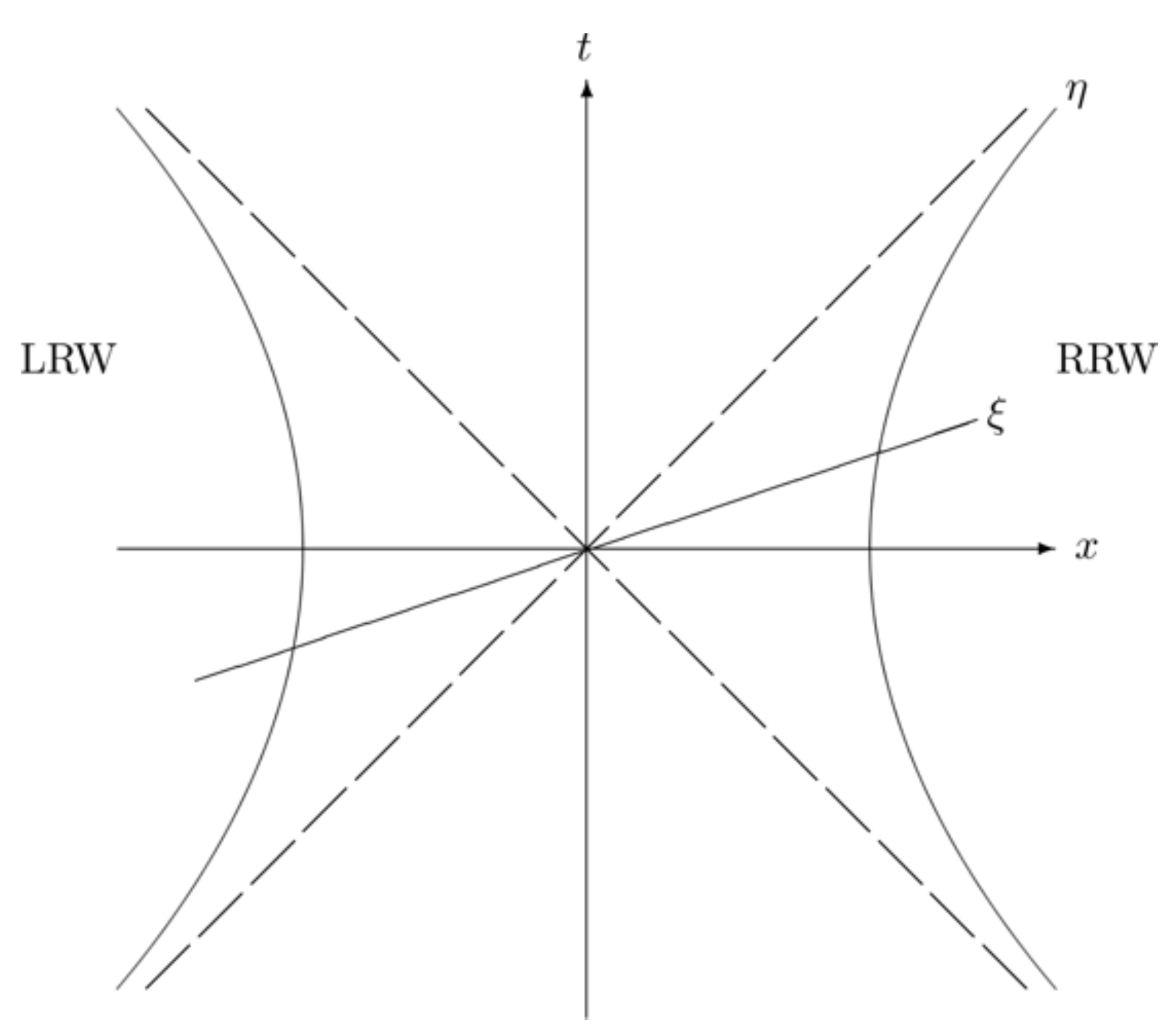}
\caption{Space-time diagram showing Minkowski coordinates $(t,x)$. RRW labels the right Rindler wedge and LRW the left. The timelike line labeled $\eta$ is a line of constant position in Rindler coordinates and is the trajectory of a uniformly accelerating observer in flat space-time. The space-like line labeled $\xi$ is a line of constant time in Rindler coordinates. The dashed lines represent the causal horizons experienced by the uniformly accelerating observers.
}
\label{embedwormhole}
\end{center}
\end{figure}
%
%
%\setlength{\unitlength}{0.8cm}

%\begin{figure}[hc]
%\begin{center}
%\begin{picture}(10,10)(-5,-5)

%	\put(-4.8,0){\vector(1,0){9.6}}
%	\put(5,-0.1){$x$}
%	\put(0,-4.8){\vector(0,1){9.6}}
%	\put(-0.10,5){$t$}
%	
%	\multiput(-4.5,-4.5)(0.535,0.535){17}{\line(1,1){.44}}
%	\multiput(-4.5,4.5)(0.535,-0.535){17}{\line(1,-1){.44}}
%	
%	\qbezier(4.8,4.5)(1,0)(4.8,-4.5)
%	\qbezier(-4.8,4.5)(-1,0)(-4.8,-4.5)
%	\put(4.9,4.6){$\eta$}
%	
%	\put(-4,-1.35){\line(3,1){8}}
%	\put(4.1,1.25){$\xi$}

%\end{picture}
%\end{center}
%\caption{Space-time diagram showing Minkowski coordinates $(t,x)$. $R$ labels the right Rindler wedge and $L$ the left. The timelike line labeled $\eta$ is a line of constant position in Rindler coordinates and is the trajectory of a uniformly accelerating observer in flat space-time. The space-like line labeled $\xi$ is a line of constant time in Rindler coordinates. The dashed lines represent the causal horizons experienced by the uniformly accelerating observers.}
%\end{figure}

Techniques used to study the Unruh effect have been adopted to look at quantum information processes and entanglement in the presence of uniformly accelerating observers. Early studies of an accelerating atomic qubit found that the interaction with the Unruh thermal bath caused decoherence of the state of the qubit \cite{kok:2003}. The decoherence was shown to increase with increasing acceleration, making the state completely mixed in the limit of infinite acceleration. This decoherence is, in a way, just the usual Unruh effect, that is the thermalisation of an accelerated detector, put into the language of quantum information theory. What the authors pointed out from this alternative perspective was that if the atom had been entangled with another qubit, to create a resource for performing a task like the quantum teleportation protocol, (see section 3.2 for a description of teleportation) then the decoherence would reduce the protocol's performance. That is, the ability to transfer the quantum state of an inertial system to an accelerating system would be hampered by the thermal radiation bathing the accelerated observer. 

The reduction of fidelity of teleportation due to the Unruh effect was simultaneously confirmed \cite{alsing:2003%, alsing:2004
} without the use of atoms, but with the entanglement resource being between two modes of a scalar field. It should be noted that in this case the half of the entangled state as seen by the accelerating observer is being directly thermalised by the Unruh effect, in contrast to the atom which is interacting with the thermalised field. Having calculated the teleportation fidelity, they showed that, like the purity of the atom, the fidelity decreased with increasing acceleration and would asymptote to zero in the limit of infinite acceleration. From this it was inferred that the reduction of fidelity was indeed due to loss of the original pure state entanglement between the two field modes. This intuition was quickly confirmed by a direct calculation of the entanglement between two field modes when half of the original state is seen by a uniformly accelerating observer \cite{fuentes:2005}. Like the calculation of the teleportation protocol, a maximally entangled pure state of a scalar field was investigated and once again the dependence was found that loss of entanglement increased with increasing acceleration, and the entanglement would asymptote to zero in the limit of infinite acceleration.

These initial findings regarding decoherence and entanglement loss from the Unruh effect have since been greatly expanded on. Further developing the case of entanglement between modes of a free field, investigators have studied the fermionic Dirac field and contrasted it to the original results of the bosonic scalar field \cite{alsing:2006, martinmartinez:2009, juan:2009}. The first outcome of this was the discovery that for fermionic fields in the limit of infinite acceleration a finite amount of entanglement always persisted. This generated the hope that entangled systems could be found which are protected from Unruh radiation. This possibility was further seen when the scalar field of the original works was replaced by the full electromagnetic field \cite{ling:2007}. The result was that there exists degrees of freedom in the electromagnetic field , the photon helicity, where entanglement can exist which is completely robust against acceleration radiation. Indeed, even when considering the original scalar field, if cavities are used to store the quantum information, entanglement can be both created and stored despite the Unruh effect \cite{downes:2010}. 

We have now seen that any initial number state entanglement in bosonic and fermionic fields will be reduced if one of the observers is uniformly accelerating. We have also seen entanglement is somewhat protected in fermionic fields, and even more so in the helicity of photons and when using high quality cavities. These findings have all focused on gradual entanglement loss due to the Unruh effect, however it turns out that the Unruh effect not only degrades entanglement but can exhibit both sudden death of entanglement and creation of entanglement as well. This has been seen between two accelerating atoms interacting with the Unruh bath, demonstrating both entanglement creation \cite{benatti:2004, massar:2006, zhang:2007} as well as entanglement sudden death (where entanglement completely disappears for finite accelerations) \cite{celeri:2010, wang:2010}. It might have been thought that the Unruh effect would always destroy any entanglement between the two atoms, however their interaction through the field has been shown to generate entanglement in spite of the thermal radiation. The interaction of two atoms with the scalar field when one or both is accelerating has been developed in a rigorous mathematical framework demonstrating a rich collapse and revival structure of the entanglement dynamics \cite{lin:2006, %lin:2008, 
lin:2009}.

%The creation of entanglement has also been seen between field modes when one considers the dependance on the initial state chosen \cite{montero:2010}. The effects on initial states other than the maximally entangled pure state have also been considered for example with nonmaximal entanglement \cite{pan:2008}, tripartite entanglement \cite{hwang:2011, wang:2011} and mixed states \cite{moradi:2009}. These investigations have all added to understanding of the structure of entanglement when viewed by accelerating observers. 

%In order to fully understand the dependence of quantum correlations on acceleration which go beyond just looking at entanglement, quantum discord has been studied for both the scalar and Dirac fields in non-inertial frames \cite{datta:2009, wang:2009, dehnavi:2010}. It was found that quantum correlations can still exist in regimes where it had previously been shown that entanglement would vanish.

So far we have only considered the case of acceleration which is uniform for all time. The more difficult case is non-uniform acceleration, for example, where an initially inertial observer begins to accelerate \cite{Sch04}. This particular case has been tackled including a way to track the quantum correlations in time \cite{mann:2009}. One must also deal with the problem that for non-uniform acceleration the symmetries required to even talk about the creation and annihilation of particles do not exist globally. The solution is to consider the particle states and entanglement on space-like hypersurfaces where the required symmetries do exist. The result is that when acceleration is not uniform for all time, entanglement reduces to a finite amount even for free bosonic fields.  

\subsection{Entanglement in the vacuum} 

In Minkowski space-time entanglement exists between different, even space-like separated, regions of the quantum vacuum. This can be seen directly by expanding the Minkowski vacuum, $|vac \rangle_M$, in terms of the Rindler modes - wave equation solutions in terms of the Rindler coordinates (Eq.\ref{R1} and \ref{R2}). One obtains:
\begin{eqnarray}
|vac \rangle_M = \Pi_i \:\Sigma_{n_i=0}^\infty \; c_i \; e^{-\pi n_i \omega_i/a} |n_{\omega_i} \rangle_R |n_{\omega_i} \rangle_L
\label{E1}
\end{eqnarray}
where $|n_{\omega_i} \rangle_R$ is the state of a Rindler mode restricted to the right wedge, containing $n$ exitations of frequency $\omega_i$. Similarly  $|n_{\omega_i} \rangle_L$ is a Rindler mode restricted to the left wedge (see Fig.2). The $c_i$'s are normalization constants. 
This state is entangled as it is non-separable between the left and right wedges.

At the level of algebraic quantum field theory this effect arises as the mathematically rigorous result that all states of bounded energy exhibit entanglement between any two space-like separated regions \cite{halvorson:2000}. This rather technical result was demonstrated in a more physically intuitive way when it was found that two stationary, causally disconnected probes could become entangled by interacting with the field even if their state was initially separable \cite{reznik:2003}. This result also demonstrated the distance dependence of this entanglement, as the probes were further separated their entanglement would eventually vanish. In particular, if the separation between the probes was $L$ and the probes interacted with the field for a time $T$, it was found that the entanglement would disappear once $L/T > 1.1$. An analogous, discrete model that was studied is a chain of trapped ions \cite{reznik:2005}. Here, the ``vacuum state" can be defined as the collective ground-state of the normal modes of the system - a product state of the different modes. However, if this ``vacuum state" is expressed in terms of the local single oscillator states it takes the form of an entangled state. The purpose of this analogy is to illustrate the physics in a simpler, more experimentally accessible system.

%The rate of this decay of entanglement with separation was further studied and compared to the entanglement of a discrete model of a chain of atoms, where it was known that entanglement would vanish at a finite separation \cite{reznik:2004}. 
The model of two atomic probes interacting with the vacuum field was used to demonstrate violation of the Bell inequality, showing that a local hidden variable model could not account for the correlations which exist in the vacuum state \cite{reznik:2005}. It has also been noted that the Feynman propagator can have non-zero values outside the forward light cone and, although this cannot lead to communication faster than the speed of light \cite{cliche:2009}, this has led to an alternative perspective involving the propagation of virtual photons, that has been compared to the standard extraction of vacuum entanglement \cite{franson:2008}. 

The use of detectors to probe the field for vacuum entanglement has been put in a rigorous relativistic framework \cite{lin:2009}. This has demonstrated the existence of a rich dynamical structure for the entanglement between two stationary probes interacting with the field and separated by some finite distance. This entanglement dynamics is complementary to similar work on entanglement between accelerated detectors discussed in the previous subsection.      

Another possibility is to probe for entanglement between timelike separated regions of space-time. Indeed, it has been found that for massless fields there exists entanglement between the past and the future in the quantum vacuum \cite{OLS11}. Surprisingly, a thermal response exactly analogous to that of an accelerated detector arises for a stationary detector with suitably time-varying energy levels. In fact a suitably scaled detector couples to precisely the same Rindler modes (see Eq.\ref{E1}) as those for an accelerated detector. Calculations suggest that the effect could be visible for energy gaps on the order of 100 GHz, scaled over a single order of magnitude in a pico-second. It has further been shown that two suitably scaled detectors, one interacting with the vacuum in the past, and the other only interacting with the future vacuum, will become entangled, thus extracting the timelike vacuum entanglement for potential applications \cite{OLS11a}.

\subsection{Black-holes}

A black-hole is a region of space-time surrounded by an event horizon due to an extremely strong gravitational field, i.e. strong space-time curvature \cite{taylor:2000}.  Space-time curvature in general relativity can be characterised by its metric. Flat, or Minkowski space has the metric:
\begin{eqnarray}
ds^2 = dt^2 - dr^2 - r^2(d\theta^2+\sin^2\theta d\phi^2)
\label{M1}
\end{eqnarray}
where $dt$ is an infinitesimal time interval and we have used spherical spatial coordinates and units in which $c=1$. Here $ds$ describes an infinitesimal proper distance in space-time as measured by a local observer. The proper distance between two space-time points is an invariant quantity, which all observers can agree on, regardless of the coordinate system they use. In general relativity a particle moving only under the influence of gravity will follow a geodesic of the space-time, which is the shortest path, as determined by the metric. The metric tensor can be determined for a particular matter distribution by solving the Einstein field equations. This requires knowledge of the stress-energy-momentum tensor $T^{\mu\nu}$, which describes the distribution of energy and matter. It is in this way that matter affects the curvature of space-time which in turn produces the gravitational force.

The simplest example of a black-hole is described by the Schwarzschild metric \cite{wald:1994}:
\begin{eqnarray}
ds^2&=&\left(1-\frac{2M}{r}\right)dt^2-\nonumber\\
&& \left(1-\frac{2M}{r}\right)^{-1}dr^2-r^2(d\theta^2+\sin^2\theta d\phi^2).
\label{M2}
\end{eqnarray}
where now $dt$ is an infinitesimal time interval as measured by an observer far from the black-hole, $r = circumference/2 \pi$ is the reduced circumference and the other spherical coordinates remain as for flat space. We are using units now where $G$, the universal gravitational constant and $c$, the speed of light, are both set to unity. $M$ is the mass of the black-hole expressed in geometric units (metres). Time-like proper intervals for stationary observers ($dr = d\theta = d\phi = 0$) close to the black-hole appear longer to far away observers, i.e. clocks run slower. As we approach $r = 2 M$ clocks appear to stop. This is the event horizon. 

Traditionally anything which passes through the horizon, including light, can never escape the immense gravitational field. When one considers quantum fields propagating on a black-hole space-time however, it is found that a black-hole should emit a thermal distribution of particles at the Hawking temperature \cite{hawking:1975}:
\begin{eqnarray}
T_H &=& \frac{\hbar c^3}{8\pi GMk}.
\label{H}
\end{eqnarray}   
This is in direct analogy to the Unruh effect, indeed the Unruh effect was discovered as part of a study of the Hawking effect. That is, the section of the quantum vacuum trapped behind the event horizon is entangled with that outside. Observers outside the black-hole have no access to the field inside and so see a mixed, thermal state.
%A rough way of seeing this is through the equivalence principle. Consider a blackbody suspended at radius $r$ from the black-hole. It will feel an acceleration due to the gravitational field of the black-hole $a = $ and hence have an Unruh temperature of $T= $ (Eq.\ref{U}). The blackbody radiation from the object, as observed far away from the black-hole, will be red-shifted by a factor $\sqrt{(1-\frac{2M}{r})}$. Thus to a far away observer the temperature of the blackbody will appear to be $T= $. If we now ask what temperature a blackbody suspened at the horizon, ($r= 2 M$, see below) will have we obtain the Hawking temperature (Eq.\ref{H}).

The implication of this thermal radiation is that any information entering the black-hole will be irreversible lost. When the time comes that the black-hole evaporates entirely, all that will remain is thermal radiation. This was thought to have serious ramifications for the unitarity of quantum mechanics and so, much effort has been dedicated to resolving it. Quantum information theory has been used by authors to explore various mechanisms via which the information might be returned \cite{livine:2006, lloyd:2006, smolin:2006}. Others have alternatively argued that information can never be fully recovered when lost into black-holes \cite{braunstein:2007} and that information loss may actually be an intrinsic phenomena \cite{gambini:2004} and not actually a problem for quantum mechanics after all \cite{unruh:1995}.  

Just like for the Unruh effect, quantum entanglement has also been investigated between two relatively moving observers in a black-hole space time. It has been found that two initially entangled observers will lose their entanglement if one of them moves close to a black-hole \cite{matinez:2010, pan:2008b}. This loss of entanglement is due to the thermal radiation from the Hawking effect and is in direct analogy to entanglement loss due to the Unruh effect. However, just like in the Unruh effect, entanglement can also be created between an observer near a black-hole and one far away with the use of projective measurements \cite{han:2008, wang:2010b}. Entanglement between modes inside and outside a black-hole have been shown to emerge in the formation of black-holes \cite{martinez:2010b}. This entanglement found in black-hole physics is believed by some to play a key role in recovering the unitary of Hawking radiation.

Within the semi-classical approximation of black-hole evaporation two main arguments exist for recovering the unitarity of quantum mechanics. The first is that the information is encoded in correlations. Here one may consider an analogy to classical information theory. An encrypted message gives no information without the key, and the key by itself does not contain the message. It is the correlations between the two which reveal the secret information. It is in this way that information lost in the form of Hawking radiation may be recovered in the entanglement which exists in the field \cite{smolin:2006, nikolic:2009, belokolos:2009}. However others argue that not all the information can be accounted for by the correlations \cite{braunstein:2007}. The second major argument for unitarity involves the idea of black-hole remnants and final state projection models. The semi-classical approximation breaks down in the final moments of black-hole evaporation and so it is conjectured that instead of completely evaporating a remnant might remain which contains the missing information \cite{banks:1993}. This is similar to the final state projection models where the solution is to place boundary conditions on the final state of the black-hole in order to recover unitarity \cite{horowitz:2004, lloyd:2006}.

The other side of the debate is that information is intrinsically lost in black-hole evaporation and that loss of information is actually a feature of quantum mechanics not a contradiction to it \cite{gambini:2004, unruh:1995}. Intrinsic information loss due to gravity has been proposed in more general situations than black-holes and may contribute to better understanding of the quantum measurement problem and the quantum to classical transition \cite{milburn:2006}.    

Most likely the full story of black-holes and quantum information will not be fully understood until quantum gravity is fully incorporated into the picture \cite{ashtekar:2006}. Models of two dimensional quantum gravity have shown mixed results but none the less bring us closer to the complete picture \cite{fiola:1994, ashtekar:2008}. Hawking himself has even proposed his own solution using a form of quantum gravity \cite{hawking:2005}. It may also be possible that better understanding of information loss may be found by studying models of supposed non-unitarity not due to black-holes but closed timelike curves \cite{ralph:2007} - the subject of our next section.

\section{Time Machines}
\subsection{Time Machines in General Relativity}

General relativity has solutions that allow time travel. A famous example is the G\"odel Universe \cite{GOD49}. Technically these are referred to as closed timelike curves (CTC). They allow an object to follow a trajectory into its own past and hence are effectively time machines. Standard relativistic quantum field theory is incompatible with such metrics \cite{birrell:1982}.

A controversial question is whether CTCs can be brought into existence within a space-time like our own universe. On the one hand schemes have been proposed whereby CTCs might be created \cite{morris:1988, GOT91}. In turn such schemes have been criticised on various grounds \cite{DES92, HAW92}. So far the argument appears undecided \cite{KIM91, VIS03, ORI05, EAR09}. Essentially all such discussions require assumptions to be made about the nature of the so far undiscovered complete theory of quantum gravity. The outcome - CTCs or no CTCs - seems to depend on these assumptions.

A specific example of a general relativistic time machine, potentially compatible with our own universe, is the traversable wormhole metric introduced by Morris and Thorne \cite{MOR88A}. Whilst wormholes, bridges between different parts of our universe, have been known to be solutions to Einstein's field equations  for many years \cite{FLA16, EIN35}, they were also known to be non-traversable for several reasons including: their extremely rapid time evolution; and the presence of a horizon within the wormhole throat. In contrast, Morris and Thorne's solutions are static and contain no horizons. %(see also ?). 
Consider the metric:
\begin{equation}
ds^2 = -e^{2 \Phi} dt^2 + dr^2 {{1}\over{1-b(r)/r}} + r^2(d \theta^2+sin^2\theta d \phi^2)
\label{wh}
\end{equation}
Here $t$ is the time coordinate of a static observer, $r$ is the radial coordinate, and $\theta$ and $\phi$ are again polar coordinates. $b(r)$ is the shape function of the wormhole and $\Phi(r)$ is the red-shift function. A particularly benign solution arises if we take $\Phi(r)=0$ and $b(r) = b_0 (1-(r-b_0)/a_0)^2$ for $b_0 < r< b_0+a_0$, and $b(r) = 0$ for $r>b_0+a_0$. An observer traveling radially toward the wormhole begins in flat space with a monotonically decreasing radial coordinate. The radial coordinate stops being monotonic when $r=b_0+a_0$; it decreases to a minimum radius of $b_0$, the wormhole throat, at which point it starts to increase again until we ``emerge" again into a different region of flat space when $r$ is again larger than $b_0+a_0$. It can be shown that such a radially launched observer would travel at a constant velocity, $v$, through the wormhole. Furthermore, if this velocity obeys $(v/c)^2 \lesssim a_0 b_0 /(10^8 metres)^2$ then the traveller will encounter negligible tidal forces. Assuming $v \ll c$, both the traveller and static observers would agree that the journey would take of order
\begin{equation}
\tau \simeq \pi a_0/v \gtrsim \sqrt{a_0/b_0}\;\; seconds. 
\label{time}
\end{equation}
%
%Fig.\ref{wormhole1} is an embedding diagram representing the wormhole in a fictitious higher dimension. 
Einstein's equations can be solved for this metric to determine what matter energy distribution is required to produce the wormhole. One finds an extreme, negative matter-energy density is required within the throat of the wormhole. Whilst prohibited classically, quantum fields can exhibit negative energy densities, though it is unknown if they can do so on macroscopic scales.
\begin{figure}[htb]
\begin{center}
\includegraphics*[width=8cm]{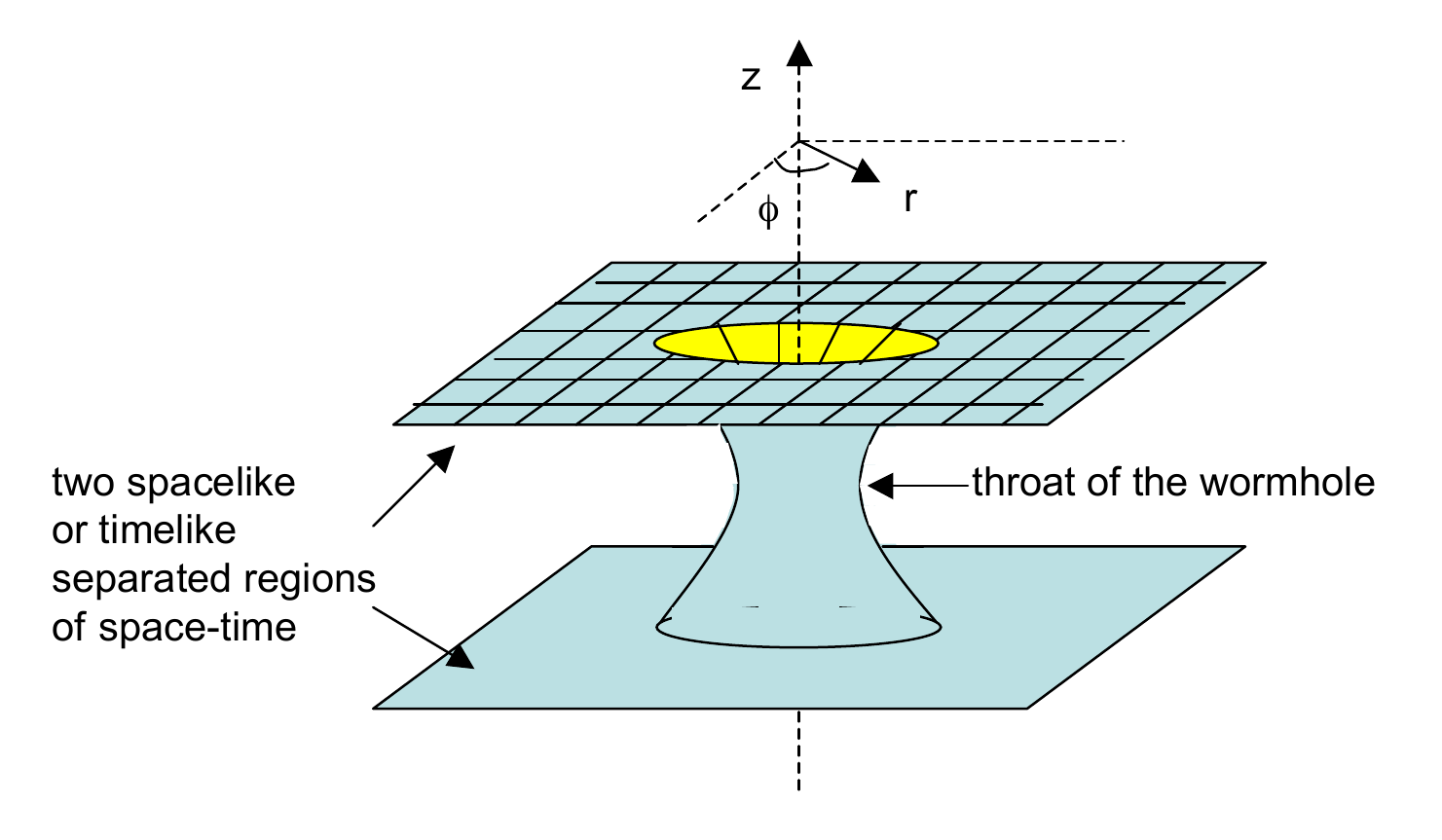}
\caption{Representation of wormhole metric via embedding diagram into fictitous higher dimension with $\theta$ and $t$ held constant.
}
\label{embedwormhole}
\end{center}
\end{figure}

Potentially, the region of space in which our traveller's radial descent begins is spacelike separated from the region in which their radial ascent ends, enabling faster than light transit between the two regions. Equally well we could consider the situation in which the wormhole connects timelike separated regions, such that our traveller emerges a short distance from where they entered but at an earlier time, thus creating a CTC,  i.e. a time machine.  Morris, Thorne and Yurtsever \cite{morris:1988} gave a recipe for converting a spacelike wormhole into a timelike wormhole. They suggest that an advanced technology could drag one mouth of the wormhole on a round-trip at relativistic velocities.  Upon return, the accelerated mouth will have ``aged" less than the mouth that stayed still, creating a connection between two different times.

As we have noted, there has been considerable discussion about the validity of the CTC solution presented in the preceding paragraphs, mainly focusing on the stability of the solutions, but not leading to a clear conclusion either way. However, an alternative approach to deciding if CTCs can exist or not is to assume that they do (i.e. assume, for example, that the Morris-Thorne metric is valid), and then examine the consequences. If the consequences are not self-consistent, one might then be justified in deciding against them. A long tradition of such arguments exist. The classic example is the grandfather paradox in which the hero uses a time machine to go to the past and kill his own grandfather, thus preventing his own existence. In classical mechanics there does not seem to be any general solution to such inconsistencies. However, surprisingly, quantum mechanics appears to offer a way to avoid these paradoxes, as we shall discuss in \ref{CTC}.

\subsection{Time Machines in Quantum Mechanics}

Certain interpretations of quantum mechanics invoke time travel in order to explain its counter intuitive results. For example Richard Feynman famously interpreted the electron anti-particle, the positron, as an electron travelling backwards in time \cite{FEY48}. Another example is the transactional interpretation of quantum mechanics in which particles receive information about their future trajectories from fields propagating into the past \cite{CRA86}. This enables one to explain delayed-choice experiments \cite{WHE78} (amongst other phenomena) in which particles seem to know the future arrangement of the experiment before it is chosen.

The key feature of quantum mechanics which suggests such interpretations is that of entanglement. Consider an entangled state of two spacelike separated qubits:
\begin{equation}
|\phi^+ \rangle = 1/\sqrt{2}(|0 \rangle_1 |1 \rangle_2 + |1 \rangle_1 |0 \rangle_2)
\end{equation}
where the indicies $1$ and $2$ label the different spatial positions of the qubits. A measurement on the first qubit which gives the result {\it zero} will immediately ``collapse" the state of the second qubit into the state {\it one}, i.e. $\langle 0|_1|\phi^+ \rangle = 1/\sqrt{2}|1 \rangle_2$. This instantaneous collapse between spacelike separated regions of spacetime would seem to require faster than light influences \footnote{Strictly, we must either give up locality {\it or} realism. By realism we mean the notion that particles have objective properties before measurement.}. As we saw in the last section, when faster than light connections between spacelike separated points appear, time travel may not be too far away.

\begin{figure}[htb]
\begin{center}
\includegraphics*[width=7.5cm]{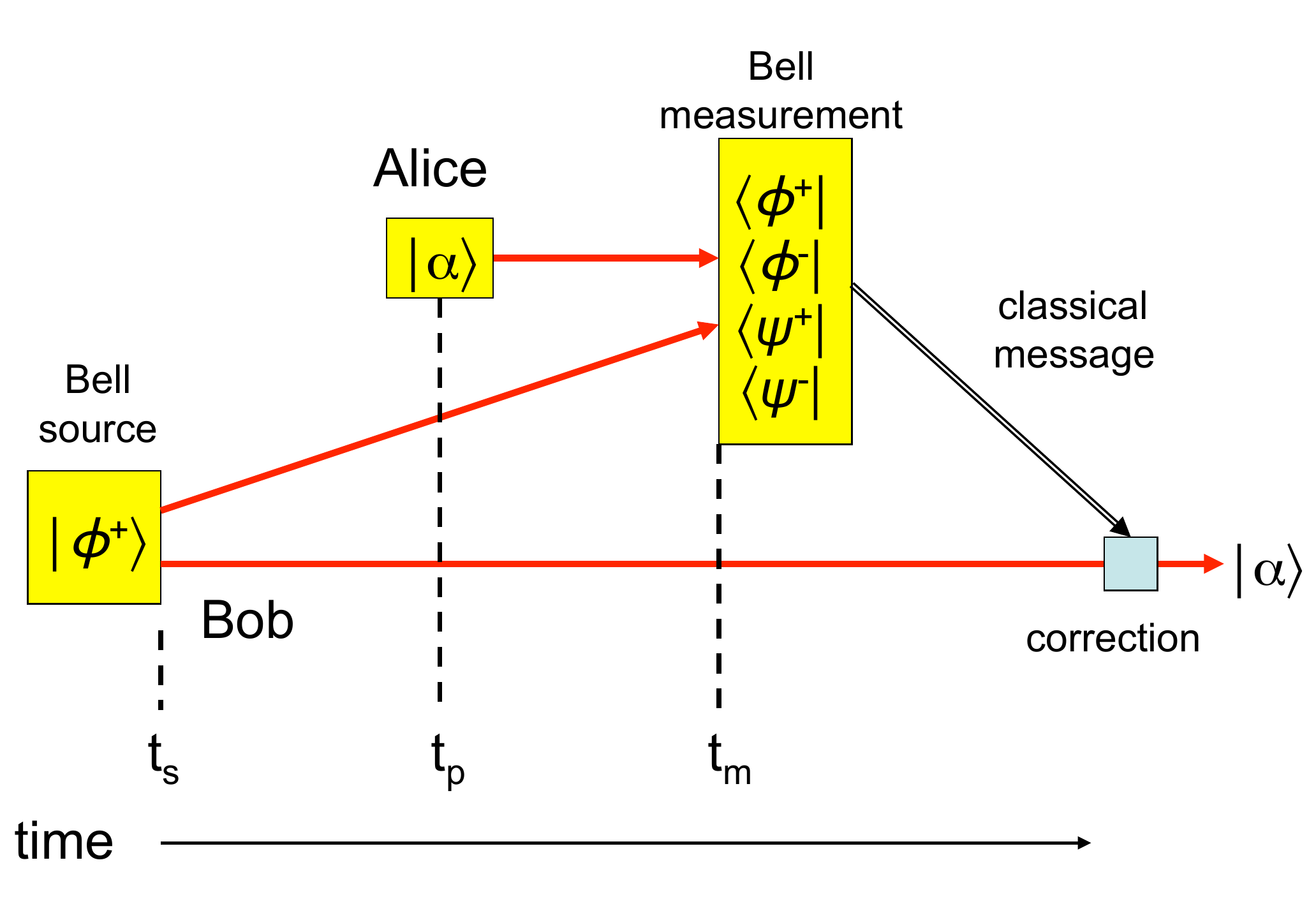}
\caption{Teleportation of a qubit. In the standard interpretation Bob keeps one arm of the Bell state, $|\phi^+ \rangle$, created at time $t_s$. The other arm he sends to Alice. Alice creates qubit state $|\alpha \rangle$ at time $t_p$ and then makes a two-mode Bell measurement between it and the arm of the entangled state received from Bob. Depending on the four possible outcomes of the Bell measurement, Bob's state collapses into a state related to Alice's state  by simple bit-flip or phase-flip operations. When Bob receives a classical message from Alice telling him which result was received, he can make the neccessary correction and obtain Alice's state. A special case is when the Bell measurement projects onto the same entangled state that Bob produced as then no correction is required. In this case one can retrodict that Bob had (a time-retarded version of) Alice's state {\it before} she created it.
}
\label{tele}
\end{center}
\end{figure}

More explicitly, it has been suggested that quantum teleportation  represents a kind of post selected time machine \cite{PEG01, BEN05, SVE09}. In the standard interpretation of teleportation \cite{BEN93} (see Fig.\ref{tele}), the sharing of entanglement between two parties, Alice and Bob, allows a quantum state held by Alice to be sent to Bob via only classical communication. In particular, Alice makes a Bell measurement on the state she wishes to send and one half of the entangled pair. The Bell measurement projects onto one of the four Bell states: $1/\sqrt{2}(|0 \rangle_1 |0 \rangle_2 + |1 \rangle_1 |1 \rangle_2); 1/\sqrt{2}(|0 \rangle_1 |0 \rangle_2 - |1 \rangle_1 |1 \rangle_2); 1/\sqrt{2}(|0 \rangle_1 |1 \rangle_2 + |1 \rangle_1 |0 \rangle_2); 1/\sqrt{2}(|0 \rangle_1 |1 \rangle_2 - |1 \rangle_1 |0 \rangle_2)$. Depending on which of the Bell states is measured by Alice, Bob's state collapses into the state Alice is sending, modulo bit and/or phase flip corrections \footnote{A bit-flip takes $|0 \rangle \to |1 \rangle$ and $|1 \rangle \to |0 \rangle$. A phase flip takes $|0 \rangle \to |0 \rangle$ and $|1 \rangle \to -|1 \rangle$.}  If Alice sends the result of her Bell measurement to Bob as a classical message then Bob can retrieve the original state by making the required corrections. An interesting case is when Alice measures the same Bell state as was originally prepared (say $|\phi^+ \rangle$). Then no correction is necessary and Bob's state instantaneously collapses into Alice's state. Of course Bob doesn't know that he already has her state until the message arrives from Alice telling him to leave the state as it is.

The interpretation of teleportation as a post-selected time machine relies on retrodictive quantum mechanics \cite{AHA64}. In particular, following Pegg  \cite{PEG01}, we retrodict the state created at the entanglement source given that $|\phi^+ \rangle$ was measured and the state $|\alpha \rangle$ was prepared. In order to be able to track the time evolution of the system we consider qubits constructed from energy eigenstates, such that the initial qubit created, 
\begin{equation}
|\alpha \rangle= \mu |0 \rangle + \nu |1 \rangle
\label{qa}
\end{equation}
evolves to $|\alpha \rangle= \mu |0 \rangle + e^{-i \omega \tau} \nu |1 \rangle$ after a time interval $\tau$. Suppose the qubit state is prepared at time $t_p$, whilst the Bell pair is created at an earlier time $t_s$ and the  Bell state measurement is at a later time $t_m$ (i.e. $t_s < t_p < t_m$, see Fig.\ref{tele}). We start from the Bell measurement and evolve the projected state, $|\phi^+ \rangle$, back to the preparation at $t_p$ (note that $|\phi^+ \rangle$ is an energy eigenstate and so acquires only a global phase from evolution). Preparation of the qubit state is modelled as a projection onto the retrodicted entangled state \cite{PEG01}: 
\begin{equation}
\langle \alpha|\phi^+ \rangle = \mu^* |1 \rangle + \nu^* |0 \rangle.  
\end{equation}
This state is evolved back in time to the source to give us $\mu^* e^{-i \omega(t_s-t_p)}|1 \rangle + \nu^* |0 \rangle$. Finally we project this state onto the state prepared at the Bell source to give: 
\begin{equation}
 (\mu e^{i(t_s-t_p)\omega} \langle 1| + \nu  \langle 0|)|\psi^+ \rangle =  \mu e^{i(t_s-t_p)\omega} |0 \rangle + \nu |1 \rangle. 
 \label{qb}
\end{equation}
This state is the time retarded version of Alice's qubit (Eq.\ref{qa}) but appearing in Bob's mode {\it before} it was created by Alice. This is the sense in which teleportation is a time machine. If we propagate Bob's state forward in time to $t_p$ we obtain (up to a global phase) the state that Alice prepares \footnote{That is, propagating the state of Eq.\ref{qb} forward in time by the interval $t_p-t_s$ gives us the state $e^{i(t_s-t_p)\omega} ( \mu |0 \rangle + \nu |1 \rangle)$ which is Alice's initial state. The (global) phase factor multiplying the entire state has no physical significance.}. If we propagate to the time of the Bell measurement, $t_s$, then Bob's state is:
\begin{equation}
|\alpha \rangle= \mu |0 \rangle + e^{-i (t_m-t_p)\omega} \nu |1 \rangle
\end{equation}
which is the same as the state Alice teleports in the standard approach - showing the two pictures coincide after time $t_m$. Of course all of this only works when the state $|\phi^+ \rangle$ ends up being the one measured, and that only happens one quarter of the time. Overall, in the absence of knowledge about the eventual result of the Bell measurement, Bob holds a thermal state at time $t_m$.

If we take the time machine interpretation of teleportation seriously we might ask what happens if we try to cause a paradox by giving Alice an inconsistent qubit to teleport \cite{PEG01}. For example, Bob might take the qubit he receives at time $t_p$ and bit flip it, before giving it to Alice and telling her to teleport it to him in the past. The bit flipped qubit is orthogonal to the one he actually received and so creates an inconsistent history. The resolution of the paradox is straightforward - the probability of obtaining the measurement result corresponding to $|\phi^+ \rangle$ is zero, i.e. the time machine never works. Some authors have suggested that this can form the basis for avoiding paradoxes in CTC's by renormalizing to only the self consistent solutions of the circuit, as we will discuss in the next section.

%\subsection{Teleportation to the Past?}

%Stuff

\subsection{Quantum Mechanics on Closed Timelike Curves} \label{CTC}    

\begin{figure}[htb]
\begin{center}
\includegraphics*[width=7.5cm]{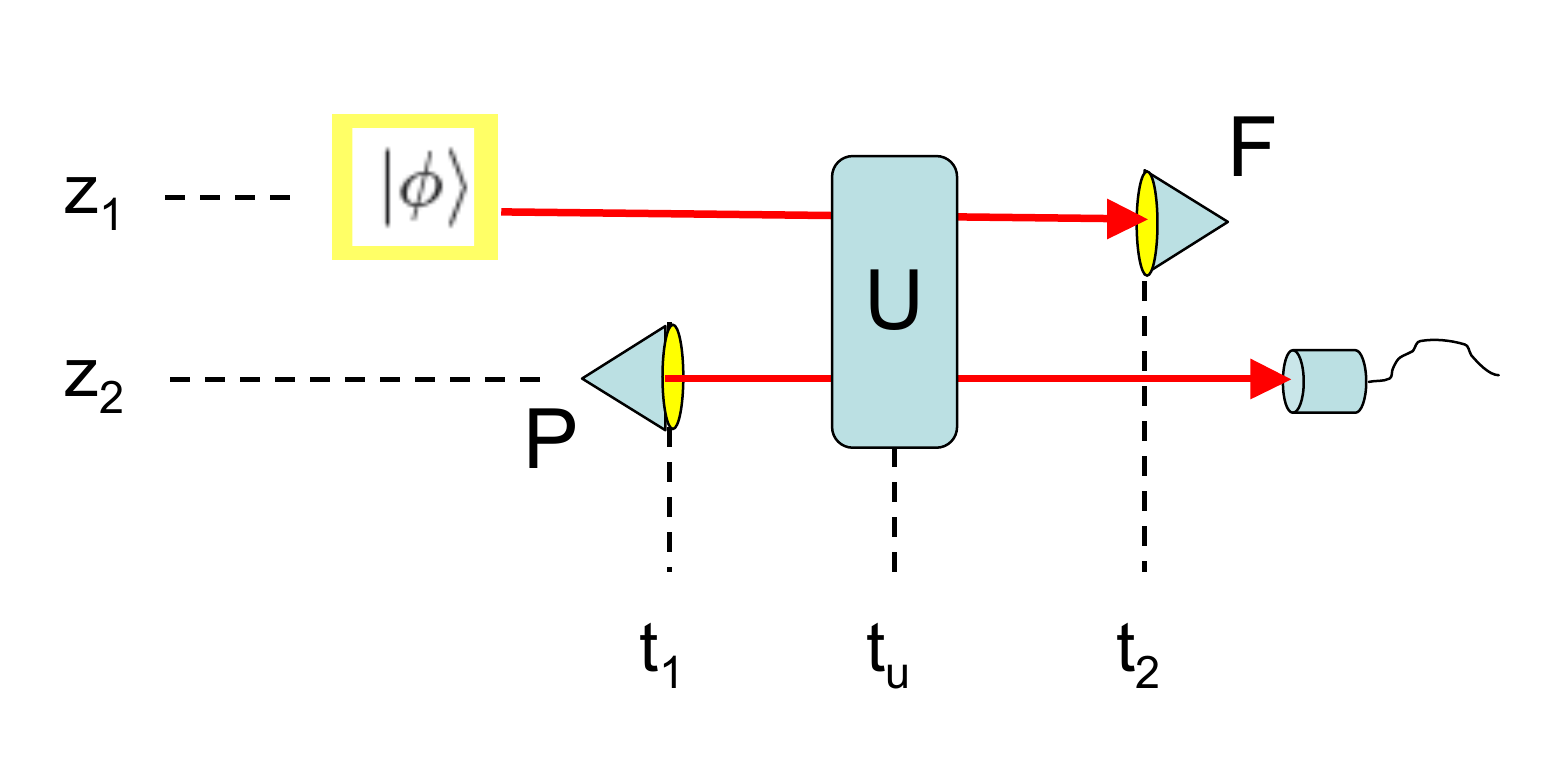}
\caption{Model of a qubit interacting with a closed timelike curve formed by a wormhole, as seen by an observer far away from the wormhole. The qubit suffers an elastic collision with a second qubit described by the unitary $U$. The qubit can be scattered into the future mouth of the wormhole (F). It emerges from the past mouth of the wormhole (P), in the past, and becomes the second qubit in the elastic collision, thus forming a closed timelike curve. It can then be scattered into an output mode that is subsequently detected.
}
\label{wormhole}
\end{center}
\end{figure}

%Although the existence of closed timelike curves (CTCs) Ð essentially time machines Ð is highly speculative, there are a number of reasons for studying how quantum system might interact with them:
%(i) CTCs are part of the mathematical structure of General Relativity. Standard quantum field theory is incompatible with CTCs. Investigating modifications of field theory that are compatible with CTCs might shed light on other inconsistencies between quantum mechanics and relativity.
%(ii) Surprisingly, toy models of quantum systems appear to be more compatible with CTCs than classical systems. Understanding exactly why this is so could shed light on unresolved issues at the quantum-classical boundary.
%(iii) The toy models exhibit fundamentally non-linear behaviour, very different to standard quantum mechanics, and breaking basic tenets such as the uncertainty principle. It is of considerable interest to investigate whether the toy models can be consistently extended to general systems Ð realizing a long sought after non-linear extension of quantum mechanics. Such an extension may exhibit testable features in low curvature situations.
In this section we will examine what happens if we assume wormhole/time-machine metrics, such as Eq.\ref{wh}, are valid descriptions of possible structures in our own universe and place simple quantum systems through them, allowing them to interact with their own past incarnations. The model usually considered can be portrayed as in Fig.\ref{wormhole}. A two-level quantum system, qubit 1, suffers an elastic collision, i.e. unitary interaction, with qubit 2, and then enters a wormhole-time-machine which brings it out in the past, where it finds that in fact it {\it is} qubit 2. The qubit then propagates away from the worm-hole where it can be measured. For simplicity it is assumed that the time taken to traverse the wormhole is negligible (e.g. $a_0<<b_0$ in Eq.\ref{time}) The question is whether we can create a consistent history for all input states and all unitaries, even those that at first sight appear paradoxical.

Two main approaches have been discussed in the literature. The first we will consider was suggested by Politzer \cite{POL94} (see also \cite{ECH91}). The standard Feynman path-integral approach of quantum mechanics is applied to the circuit to work out the probabilities of all particular outcomes given the boundary condition that the field at $(z_1, t_2)$ matches that at $(z_2, t_1)$ where $t_1<t_2$ (see Fig.\ref{wormhole}) for each possible history. In general, unlike the case for standard quantum mechanics, the sum of all these probabilities will not add up to one. Thus to preserve the probabilistic interpretation of quantum mechanics, the final state must be renormalised. For example, suppose that the initial state is $\alpha |0 \rangle + \beta |1 \rangle$ and the unitary is a controlled-NOT (CNOT) with the upper rail as the control and the lower as the target. In a CNOT gate the logical state of the target qubit is flipped if and only if the control qubit is in the logical {\it one} state. For simplicity here we consider stationary qubits whose free evolution is the identity. We first calculate the trajectory of the {\it zero} component of the state. A logical state traversing a control is unchanged, so we can immediately say that the only trajectory consistent with the boundary conditions is
\begin{equation}
|0 \rangle \to |0 \rangle|0 \rangle \to |0 \rangle|0 \rangle \to |0 \rangle
\end{equation}
where we are moving progressively through time intervals $t < t_1$ to $t_1< t < t_u$ to $t_u< t < t_2$ to $t > t_2$. Following the same procedure for the {\it one} component of the state we find
\begin{equation}
|1 \rangle \to |1 \rangle|1 \rangle \to |1 \rangle|0 \rangle \to |0 \rangle.
\end{equation}
The output state is formed from the coherent superposition of the two paths: $1/\sqrt{2}(\alpha+\beta)|0 \rangle$. Renormalizing we find that the output is always the ground state, $|0 \rangle$, regardless of the input state. All paradoxical or inconsistent histories are suppressed. However, the renormalization procedure is state dependent, which means the linearity of quantum mechanics is lost. More worryingly, state dependent renormalization means that events in the far future can effect experiments performed now, as we will discuss in the next section. Interestingly, as noted at the end of the last section, this procedure turns out to be equivalent to renormalising the outcome of an equivalent teleportation time machine circuit to the $|\phi^+ \rangle$ outcome \cite{LLO11}. 

The second approach we consider is due to Deutsch \cite{DEU91}. Instead of requiring individual components of the state vector to satisfy the boundary conditions (and suppressing them if they do not) as in the the path integral approach, Deutsch instead required that the entire state at $(z_1, t_2)$, as defined by its reduced density operator $\rho$, matches that at $(z_2, t_1)$. Surprisingly, this requirement can always be satisfied regardless of the initial state or unitary. The Deutsch solution is to firstly determine the state of qubit 2, given by the density operator $\rho$, via the consistency equation
\begin{equation}
\rho = Tr_2[U(\rho_{in} \otimes \rho)U^{\dagger}]
\label{SP1}
\end{equation}
where the trace is over the Hilbert space of qubit 2. Given a solution for $\rho$, the output state of qubit 2, $\rho_{out}$, is given by
\begin{equation}
\rho_{out} = Tr_1[U(\rho_{in} \otimes \rho)U^{\dagger}]
\label{SP2}
\end{equation}
where the trace is now over the Hilbert space of qubit 1. In general the solution for $\rho_{out}$ will be a non-unitary and a nonlinear function of the input state. However, the character of the transformation produced is quite different from the path integral approach. Consider again the example of a pure input state and a CNOT gate. To apply the Deutsch method we must first identify the density operator, $\rho$, which satisfies the consistency requirement Eq. \ref{SP1}. It is straightforward to confirm that  
\begin{equation}
\rho =  |\alpha|^2 |0 \rangle \langle 0| +  |\beta|^2 |1 \rangle \langle 1|
\label{1}
\end{equation}
satisfies Eq. \ref{SP1}. Substituting $\rho$ into Eq.\ref{SP2} results in the output state
\begin{equation}
\rho_{out}  =  (|\alpha|^4 |+ |\beta|^4 |) |0 \rangle \langle 0| + 2 |\alpha \beta|^2 |1 \rangle \langle 1|.
\label{1}
\end{equation}
The solution is clearly non-unitary as $\rho_{out}$ is diagonal and hence all coherences have been lost.  The output is also a non-linear function of the input state, however, because there is no renormalization of the input state, the future cannot affect the past outside the region of the CTC, as can happen in the path integral approach (see next section). Various authors have investigated the increased information processing power implied by CTC's of this kind. For example, using the state evolution of Eq.\ref{1}, a quantum computer could use a CTC in a subroutine to solve NP-complete problems \cite{BAC04}. Put simply, NP is a complexity class corresponding to problems that cannot be solved quickly on a classical computer, but whose solutions can be quickly checked. Some problems believed to be in NP (such as finding the prime factors of an integer) can be solved quickly on a quantum computer. A sub-class of NP, NP-complete, is comprised of NP problems that it is believed cannot be solved quickly with a quantum computer. The non-linearity introduced by the Deutsch solution of the CTC interaction changes this and allows them to be solved quickly on a quantum computer. The increased power comes from the ability of the CTC interaction to increase the distinguishability of non-orthogonal states. Indeed it has also been shown that a more sophisticated CTC circuit could be used to perfectly distinguish any finite set of non-orthogonal quantum states, in clear contravention of the uncertainty principle \cite{BRU09}.

\begin{figure}[htb]
\begin{center}
\includegraphics*[width=8cm]{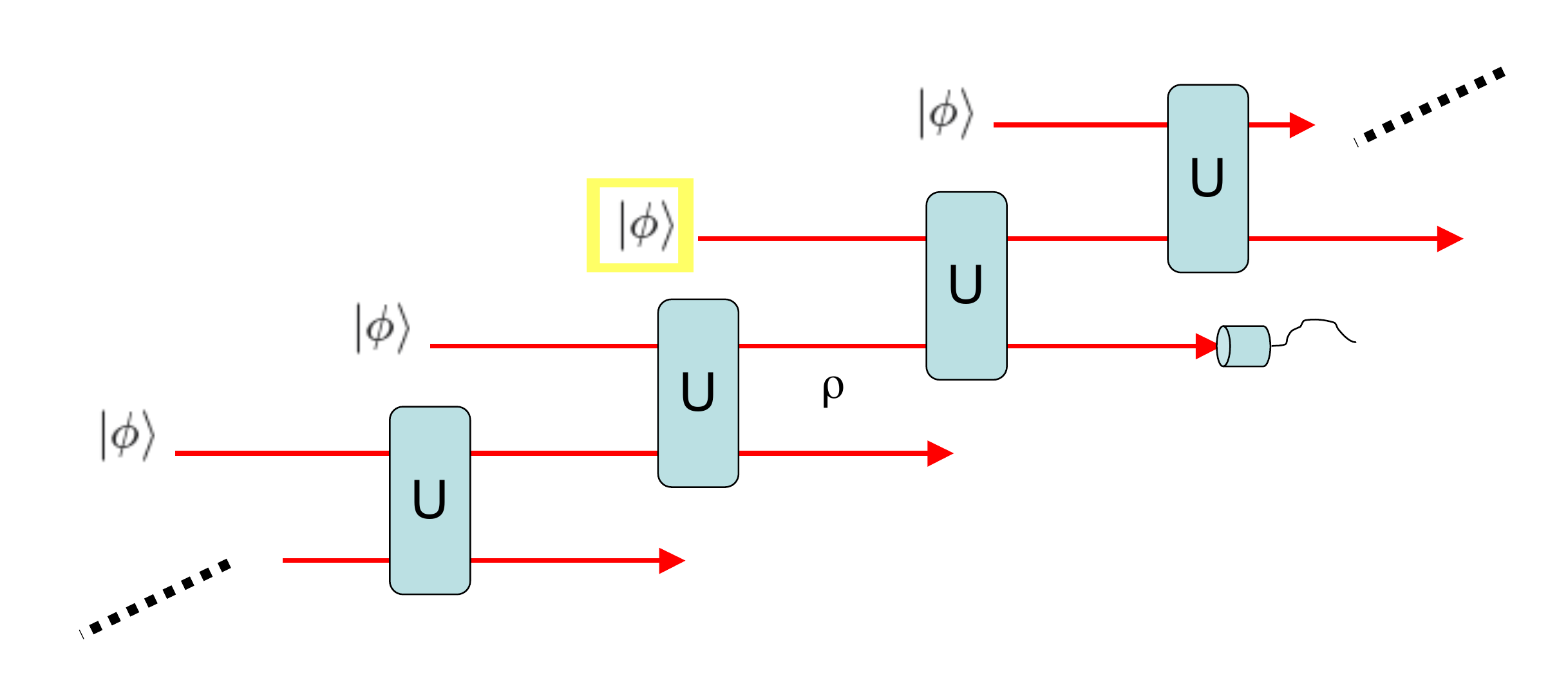}
\caption{Model of a qubit interacting with a closed timelike curve formed by a wormhole, as ``seen" by the qubit. Either propagating forward along  the path of the qubit from the initial state or back along the path of the qubit from the detector leads to infinite strings of identical interactions with copies of the qubit. Multiple interactions with the unitary represent cases where the qubit is repeatedly scattered through the wormhole. This equivalent circuit can be solved in the limit of many interactions to give a unique solution to the CTC interaction.
}
\label{EC}
\end{center} 
\end{figure}

The two approaches we have discussed both take a ``bird's-eye view" of the CTC interaction, i.e. they describe the interaction in the coordinates of an external observer who is watching from a distance. Recently Ralph and Myers \cite{RAL10} have analysed the problem in the coordinates of the qubit. Such coordinates will inevitably appear multi-valued to the far away observer due to the cycling through the CTC. In Fig.\ref{EC} the CTC interaction is expanded into an equivalent circuit. This equivalent circuit can be obtained heuristically by considering what the qubit ``sees" when propagating forward through the wormhole or alternatively, by tracing its trajectory back through the wormhole into the past. In both cases the qubit experiences multiple interactions via the unitary $U$ with copies (i.e. perfect clones) of itself. The effective circuit stretches indefinitely into the future and into the past, representing the time lines experienced by the qubit, even though, to the far away observer all these paths overlap in the same time interval. 
%-- an important point to which we will return shortly. 
%We have assumed that the action of the wormhole in creating the CTC is purely geometric, as described by general relativity. The equivalent circuit uniquely describes that geometry as viewed from the perspective of the qubit. The behaviour of the qubits on the equivalent circuit are then assumed to follow the normal rules of quantum physics.
The interpretation of the equivalent circuit is clarified by considering the limit in which there is no interaction with the wormhole. This occurs when $U=SWAP$, i.e. when the action of the unitary is simply to exchange qubits 1 and 2. In order to recover standard quantum mechanics in this limit we must assume that all the effective modes except the one indicated to be striking the detector, are lost (i.e. not measured). The other modes might be interpreted as additional degrees of freedom that are not normally observed but can be indirectly probed via the interaction $U$ in the presence of the CTC created by the wormhole.
%Calculating the Heisenberg evolution of the effective circuit of Fig.1(b) leads to the same expectation values as derived in \cite{RAL07}, where the additional degrees of freedom are associated with the space-time geometry.
%

So, what does the equivalent circuit predict? We can solve the equivalent circuit by starting far in the ``past" with the initial state for the upper arm $\rho_{in}$, and that of the lower arm the maximally mixed state $I$ \footnote{This choice is justified by considering the stability of the circuit to very small amounts of decoherence in the unitary U}. After one iteration we obtain $\rho' = Tr_L[U(\rho_{in} \otimes I )U^{\dagger}]$ where the lost (lower) arm has been traced out. After two iterations we have $\rho'' = Tr_L[U(\rho_{in} \otimes \rho')U^{\dagger}]$. If this map converges to a fixed point then it will be true, after many iterations, that $\rho = Tr_L[U(\rho_{in} \otimes \rho)U^{\dagger}]$. This expression coincides with that given by Deutsch for calculating the corresponding $\rho$  (Eq.\ref{SP1}). Deutsch showed that such a fixed point always exists \cite{DEU91}, so the assumed convergence is guaranteed. Further, given that the part of the effective circuit that propagates into the ``future" is also lost, then the output state is given by $\rho_{out} = Tr_1[U(\rho_{in} \otimes \rho)U^{\dagger}]$ where now the trace is over the upper qubit. Again this corresponds to the expression given by Deutsch for calculating the output state (Eq.\ref{SP2}). Thus the circuit of Fig.\ref{EC} is mathematically equivalent to the Deutsch solution of Fig.\ref{wormhole}. 

In addition to reproducing the Deutsch result, the equivalent circuit solution, resolves two of its ambiguities: (i) The equivalent circuit always produces a unique solution, unlike Deutsch's consistency requirement that is sometimes unconstrained. The unique solution of the equivalent circuit corresponds to Deutsch's maximum entropy conjecture \cite{DEU91}; and (ii) the ambiguity over how to treat classically mixed input states \cite{BEN09} is resolved by the equivalent circuit treatment \cite{RAL10}. 
%The EC formulation of DeutschÕs problem is the most self-consistent model of interactions of quantum systems with CTCÕs. However it remains a toy model of 2-state systems with no spatio-temporal extent. 

\subsection{Entanglement and Time Machines}

The fact that the equivalent circuit's simple geometric picture immediately reproduces Deutsch's boundary conditions suggests that the Deutsch approach may be more self-consistent than the path integral approach. Further evidence for this conclusion follows from considering the behaviour of entangled states for which one arm of the entanglement is passed through the time-machine wormhole. In particular let us suppose that the entangled state $|\phi^+ \rangle$ is created by Alice at some time $t_A$. She keeps one arm and sends the other to Bob. At some later time $t_B$ Bob encounters (or creates) a compact wormhole time-machine that enables him to send his entangled particle a short time into the past. He sends his particle through the wormhole in the manner of Fig.\ref{wormhole} with a CNOT gate as the unitary interaction. It is straight forward to calculate the state of Alice and Bob. For the case of the path integral approach one finds the renormalized state:
\begin{equation}
{{1}\over{\sqrt{2}}}(|0 \rangle + |1 \rangle)_A |0 \rangle_B
\label{2}
\end{equation}
In spite of the fact that Alice shared an entangled state, she always finds her state to be diagonal whilst Bob's qubit is always in the {\it zero} state. In the absence of the CTC, standard quantum mechanics says that if Alice is unaware of Bob's actions then she always holds a maximally mixed state, regardless of what Bob does. Even stranger, because the renormalization acts on the initial state, this implies that Alice can tell by making measurements soon after $t_A$, but long before $t_B$, whether Bob will interact with a CTC or not. In other words she can predict the future actions of Bob - or alternatively, Bob can affect Alice's past \cite{RAL11}. It seems that the attempt to resolve acausal interactions with the near past by the CTC, leads to acausal effects arbitrarily far in the past through the entanglement.

The outcome is quite different when the calculation is performed using Deutsch's boundary condition. The density operator shared by Alice and Bob after the CTC is:
\begin{equation}
I_A I_B
\label{3}
\end{equation}
Now the initial entangled state has become completely decohered, which is also outside the normal purview of quantum mechanics given that the interactions involved are all unitary. This occurs due to the traces performed in obtaining the reduced density operators in the Deutsch approach, or alternatively through the leaking of information into unobserved modes in the equivalent circuit model. However, notice that in the absence of any knowledge of Bob's actions, Alice holds a maximally mixed state as predicted by standard quantum mechanics, and in particular the actions of Bob in no way affect Alice's past. Thus, the acausal effects of the CTC do not spread outside the CTC epoch when using the Deutsch/equivalent circuit approach.

\section{Conclusion} 

In this review we have have looked at some of the key results in the new field of relativistic quantum information. We have seen how observers in different Lorenz frames can observe entanglement redistributed between the spin and momentum degrees of freedom. For non-inertial observers, entanglement can actually be destroyed by the influence of Unruh radiation. We discussed the presence of entanglement in the vacuum field, between both spacelike and timelike separated regions of space-time. Turning to curved space-time, we examined the destruction of entanglement due to Hawking radiation around a black-hole and attempts to understand the black-hole information paradox from a quantum informational perspective. 

Although much has been learned there is far more left to understand. For example, most of the investigations of entanglement have been carried out using global modes that cannot strictly be understood as objects that can be locally manipulated or measured. Only in the most recent work have techniques for modelling localized quantum systems in relativistic scenarios been developed \cite{BRU10}. These techniques need to be extended more generally. It can also be noted that discussions of Unruh and Hawking radiation generally consider free fields extending to infinity. What occurs when absorbing or reflecting boundaries exist at finite distances from the detectors? Similarly, only a few authors have considered time dependent effects. In general, more sophisticated techniques for dealing with more realistic situations are required. Ultimately we wish to find experimental quantum information systems that will exhibit relativistic effects.

In the third section we considered time-travel, firstly as it arises in the context of exotic general relativistic metrics and secondly as it arises in interpretations of quantum mechanics. We then examined two separate approaches to allowing quantum systems to interact on closed timelike curves and asked if the paradoxes of time-travel were resolved or not. Both the path integral approach and the density operator approach to matching boundary conditions on the CTC seem reasonable and appear at first to resolve paradoxical evolution, albeit in quite different ways. However, a quite distinct difference arose when entangled states were considered, whereby the path integral approach allowed a CTC created in the present to affect events in the far past, whilst the density operator approach did not. We also discussed the ``qubit-eye-view" analysis of the equivalent circuit approach, and how it agrees with Deutsch's density operator boundary condition. We might conclude that the Deutsch approach, at least for the toy models so far considered, has a good self-consistency, both internally and with the expected causal structure of the universe, and so the existence of CTCs at the quantum level cannot be ruled out on these grounds.

It is interesting to note that the path integral approach is basically standard quantum field theory with the unusual feature of initial state dependent renormalization. As we have seen this approach ultimately seems to lead to more problems than it cures - a situation which is probably linked to various CTC no-go theorems based on standard field theory. This leads one to speculate on how a field theory compatible with the Deutsch boundary condition would differ from standard theory. An initial investigation has suggested that the commutativity of space-time fields needs to be altered in order to accommodate the density operator boundary condition, and that such an alteration might have ramifications outside the realm of CTCs \cite{RAL09}. Exploring these and other ideas with more sophisticated models is likely to be a fruitful source of new understanding.

\end{document}